\documentclass{article}
\usepackage{graphicx} 
\usepackage{amsmath,amssymb,bm,mathtools,dsfont,upgreek,stmaryrd,kpfonts}
\usepackage{authblk}
\usepackage{biblatex}
\usepackage{xcolor}
\newcommand{\thmm}{\noindent {\bf Theorem. }}
\newcommand{\corr}{\noindent {\bf Corollary. }}
\newcommand{\conjj}{\noindent {\bf Conjecture. }}
\newcommand{\proof}{\noindent {\it Proof. }}
\newcommand{\qed}{\vspace{.2cm}~\hfill{$\square$}\newline\noindent}
\newcommand{\QED}{~\hfill{$\square$}}

\title{A short proof of a strong Weyl law in dimension 1}
\author[1]{August Bjerg}
\affil[1]{\centering QMATH, Department of Mathematical Sciences, \break University of Copenhagen, Universitetsparken 5, \break DK-2100, Copenhagen Ø, Denmark}
\date{}

\begin{document}

\maketitle

\begin{abstract}
\noindent For the Dirichlet realization of $-d^2/dx^2-\lambda^2V$ on a bounded interval, with $V$ a positive $C^2$ potential bounded away from $0$ and $\lambda>0$ a large parameter, we prove an asymptotic law for the values $\lambda_n$ of $\lambda$ at the $n^{\text{th}}$ appearance of a new negative eigenvalue. This approximation is correct up to an error of order $1/n$, thus making the result strictly stronger than the classical Weyl law for the number of negative eigenvalues for these operators.
\end{abstract}

\section{Main result}
Consider real numbers $a$ and $b$ satisfying $-\infty<a<b<\infty$ and a continuous potential $V\in C([a,b])$. For each $\lambda>0$, denote by $H_\lambda$ the self-adjoint Dirichlet realization of the differential operator $-d^2/dx^2-\lambda^2V$ on $L^2((a,b))$. That is, $H_\lambda$ has domain $D(H_\lambda)=\{\varphi\in H^2((a,b))\,\vert\, \varphi(a)=0=\varphi(b)\}$ inside the Sobolev space $H^2((a,b))$. We let $N(\lambda)$ denote the number of negative\footnote{By ``negative eigenvalues'' we mean always strictly negative eigenvalues, i.e. not counting $0$.} eigenvalues of $H_\lambda$.

Then the classical Weyl law for $N(\lambda)$ says that
\begin{equation}\label{Weyl}
N(\lambda)=\frac{\lambda}{\uppi}\int_a^b [V]_+^{1/2}\,dx+\varepsilon(\lambda)
\end{equation}
where $[V]_+$ is the positive part of $V$ and $\varepsilon(\lambda)$ is an error of lower order than $\lambda$ in the $\lambda\to\infty$ limit. Most standard presentations treat the case on the full space $\mathbb{R}$ and prove $\varepsilon(\lambda)/\lambda\to0$ with various assumptions on the potential (cf. \cite{RS4} Theorem XIII.79 or \cite{FrankWeyl} Theorem 4.28). Note that, since the left-hand side of (\ref{Weyl}) takes only integer values while the first term on the right-hand side grows continuously in $\lambda$, there is no chance of improving the order of the error $\varepsilon(\lambda)$ in (\ref{Weyl}) beyond constant order $\mathcal{O}_{\lambda\to\infty}(1)$.

For formulating a result which is stronger than any possible approximation of the form (\ref{Weyl}) we introduce the numbers
\[
\lambda_n:=\inf\{\lambda>0\,\vert\,N(\lambda)\geq n\}
\]
for $n=1,2,\dots$. These are the values of $\lambda$ for which $N(\lambda)$ ``jumps'' from $n-1$ to $n$. It can indeed be shown that, for all $n$, $\lambda_n<\lambda_{n+1}$ and $N(\lambda)=n$ for $\lambda\in(\lambda_n,\lambda_{n+1}]$, so that in particular $N(\lambda_n)=n-1$. Consequently, we obtain from (\ref{Weyl})
\[
\frac{\lambda_n}{\uppi}\int_a^b [V]_+^{1/2}\,dx=n+\Tilde{\varepsilon}(n)
\]
with $\Tilde{\varepsilon}(n)/n\to0$ as $n\to\infty$. Here, however, as the large parameter $n$ itself is varying discretely, there is a chance of improving the order of the error $\Tilde{\varepsilon}(n)$ beyond constant order. This is what we will do under suitable additional assumptions on the potential $V$. Our main result is the following.
\vspace{.3cm}

\thmm Assume that the potential $V$ lies in $C^2([a,b])$ and satisfies $V(x)\geq c$ for all $x\in[a,b]$ for some fixed $c>0$. Then, with the notation introduced above,
\begin{equation}\label{MainResult}
\frac{\lambda_n}{\uppi}\int_a^b V^{1/2}\,dx=n+\mathcal{O}_{n\to\infty}(1/n)
\end{equation}
where $\mathcal{O}_{n\to\infty}(1/n)$ is an error bounded by a universal constant times $1/n$.\vspace{.3cm}
\section{Proof of main result}
As an essential part of the proof of the theorem, we need the standard result of Sturm's Oscillation Theorem. That is, if $H$ is the Dirichlet realization of the operator $-d^2/dx^2+W$ on some finite interval $[\alpha,\beta]$ with a continuous potential $W$, and if $h$ solves the initial value problem $h''=Wh$, $h(\alpha)=0$, then the number of zeroes of $h$ on $(\alpha,\beta)$ equals the number of negative eigenvalues of $H$. See Section 3 of \cite{SturmOsci} for a recommended review hereof. We use this result freely below.

In the spirit of the oscillation theorem we consider a solution $f_\lambda$ to the problem $f_\lambda''=-\lambda^2Vf_\lambda$, $f_\lambda(a)=0$. Introducing the new variable
\[
\xi=\int_a^x V(y)^{1/2}\,dy
\]
and defining, for each $\lambda>0$, the function $g_\lambda(\xi)=V(x)^{1/4}f_\lambda(x)$, straightforward calculations show that $g_\lambda(0)=0$ and
\[
\frac{d^2}{d\xi^2}g_\lambda(\xi)=[-\lambda^2-U(\xi)]g_\lambda(\xi)
\]
on $(0,D)$ with
\[
U(\xi):=V(x)^{-3/4}\frac{d^2}{dx^2}\bigl(V(x)^{-1/4}\bigr)\qquad\text{and}\qquad D:=\int_a^b V^{1/2}\,dx.
\]
Note here that clearly the number of zeroes of $f_\lambda$ on $(a,b)$, i.e. $N(\lambda)$, equals the number of zeroes of $g_\lambda$ on $(0,D)$. This transformation of the equation is the cornerstone of the so-called Liouville-Green approximation, and it has a close connection to the JWKB approximation in quantum mechanics.

Observe now that from our assumptions on $V$ we can deduce the inequalities $-C\leq U(\xi)\leq C$ for all $\xi\in[0,D]$ for some constant $C>0$. Hence, we have the inequalities
\begin{equation}\label{InOfOp}
-\frac{d^2}{d\xi^2}-\lambda^2-C\leq-\frac{d^2}{d\xi^2}-\lambda^2-U\leq-\frac{d^2}{d\xi^2}-\lambda^2+C
\end{equation}
of operators acting on $L^2((0,D))$ with common domains
\[
\mathcal{D}=\{\varphi\in H^2((0,D))\,\vert\, \varphi(0)=0=\varphi(D)\}
\]
given as usual by the Dirichlet boundary conditions. By the variational principle for the eigenvalues of self-adjoint operators, cf. \cite{RS4} Theorem XIII.1, the reverse inequalities of (\ref{InOfOp}) hold for the number of negative eigenvalues of the operators. In particular, the number of negative eigenvalues of $d^2/d\xi^2-\lambda^2-U$ (which by the above equals $N(\lambda)$) lies between the number of negative eigenvalues of $d^2/d\xi^2-\lambda^2+C$ and that of $d^2/d\xi^2-\lambda^2-C$. Here, the latter are easily computed by using the oscillation theorem, and we conclude that
\[
N(\lambda)\in\biggl[\;\biggl\lceil\frac{D\sqrt{\lambda^2-C}}{\uppi}-1\biggr\rceil\,,\biggl\lceil\frac{D\sqrt{\lambda^2+C}}{\uppi}-1\biggr\rceil\;\biggr]
\]
for all $\lambda>\sqrt{C}$. By the definition of the $\lambda_n$'s this means, for all sufficiently large $n\in\mathbb{N}$, that both $n-1$ and $n$ lie in the interval\footnote{In particular, $\lambda_n\to\infty$ as $n\to\infty$ so that $\sqrt{\lambda_n^2-C-1}$ is well defined for large $n$. The fact that $\lambda_n\to\infty$ is also used below.}
\[
\biggl[\;\biggl\lceil\frac{D\sqrt{\lambda_n^2-C-1}}{\uppi}-1\biggr\rceil\,,\biggl\lceil\frac{D\sqrt{\lambda_n^2+C+1}}{\uppi}-1\biggr\rceil\;\biggr],
\]
and thus that
\[
n\in\biggl[\frac{D\sqrt{\lambda_n^2-C-1}}{\uppi},\frac{D\sqrt{\lambda_n^2+C+1}}{\uppi}\biggr].
\]
Since also $\lambda_nD/\uppi$ lies in this interval, we learn that
\[
\Bigl\vert\frac{\lambda_nD}{\uppi}-n\Bigr\vert\leq\frac{D}{\uppi}\bigl(\sqrt{\lambda_n^2+C+1}-\sqrt{\lambda_n^2-C-1}\bigr)=\mathcal{O}_{n\to\infty}(1/\lambda_n)=\mathcal{O}_{n\to\infty}(1/n)
\]
finishing the proof.\QED
\section{Perspectives}
From the discussion above it is rather clear that the classical Weyl law (\ref{Weyl}) for $N(\lambda)$ cannot by itself imply the asymptotic law (\ref{MainResult}) for $\lambda_n$. Now we argue that the converse implication holds: The above theorem implies a Weyl law for $N(\lambda)$ with an error term of order $\mathcal{O}_{\lambda\to\infty}(1)$.
\vspace{.3cm}

\corr With assumptions as in the theorem above,
\begin{equation}\label{ConstantBound}
\Bigl\vert\frac{\lambda}{\uppi}\int_{a}^{b}V(x)^{1/2}\,dx-N(\lambda)\Bigr\vert\leq 1+\mathcal{O}_{\lambda\to\infty}(1/\lambda).
\end{equation}
\proof From the theorem we learn that, for $\lambda\in(\lambda_n,\lambda_{n+1}]$,
\[
N(\lambda)=n=\frac{\lambda_n}{\uppi}\int_{a}^{b}V(x)^{1/2}\,dx+\mathcal{O}_{n\to\infty}(1/n).
\]
Consequently,
\begin{align*}
\Bigl\vert\frac{\lambda}{\uppi}\int_{a}^{b}V(x)^{1/2}\,dx-N(\lambda)\Bigr\vert&\leq\frac{\lambda}{\uppi}\int_{a}^{b}V(x)^{1/2}\,dx-\frac{\lambda_n}{\uppi}\int_{a}^{b}V(x)^{1/2}\,dx+\mathcal{O}_{n\to\infty}(1/n)
\\
&\leq\frac{\lambda_{n+1}}{\uppi}\int_{a}^{b}V(x)^{1/2}\,dx-\frac{\lambda_n}{\uppi}\int_{a}^{b}V(x)^{1/2}\,dx+\mathcal{O}_{n\to\infty}(1/n)
\\
&=(n+1)-n+\mathcal{O}_{n\to\infty}(1/n)=1+\mathcal{O}_{n\to\infty}(1/\lambda_{n+1})
\end{align*}
for all $\lambda\in(\lambda_n,\lambda_{n+1}]$. From this, the corollary clearly follows.\qed
Let us at this point notice that we believe the lack of a constant term in the right-hand side of (\ref{MainResult}) is somewhat deceptive. In fact, based on concrete examples and the analysis carried out in Chapters 2 and 5 of \cite{Thesis}\footnote{Here, Chapter 2 is a draft of \cite{PeriStructure}.} we suggest in the conjecture below a particular constant term when allowing for a more general choice of potential $V$ than in our main result. Here, the operator $H_\lambda$ is generally defined as the Friedrichs' extension of $-d^2/dx^2-\lambda^2V$ on $C_0^\infty((a,b))$, which will be bounded below for the relevant choices of potentials. In particular, Sturm's Oscillation Theorem applies to $H_\lambda$ -- see for example Theorem 4.20 in \cite{Thesis}.
\vspace{.3cm}

\conjj Suppose $V$ is sufficiently regular (say, in $C^2((a,b))$) and positive on all of $(a,b)$. If, moreover, $V$ behaves approximately\footnote{It seems likely that one would need assumptions on $V$, $V'$ and $V''$ agreeing roughly with these expressions.} like $c_a(x-a)^{\gamma_a}$ and like $c_b(b-x)^{\gamma_b}$ for some $c_a,c_b>0$ and $\gamma_a,\gamma_b>-2$ near the endpoints $a$ and $b$ of the interval respectively, then
\begin{equation}\label{Conjecture}
\vspace{.5cm}\frac{\lambda_n}{\uppi}\int_a^b V^{1/2}\,dx=n+\frac{1}{4+2\gamma_a}+\frac{1}{4+2\gamma_b}-\frac{1}{2}+o_{n\to\infty}(1).
\end{equation}
Here, the assumptions in the above theorem corresponds to taking $\gamma_a=0=\gamma_b$. It is not unreasonable to believe that the conjecture can be proved following essentially the strategy of the proof of the main result of this note, although one would need to tackle many more technical issues in the more general case. Some examples of potentials for which the conjecture can be straightforwardly verified are $V(x)=(x-a)^{\gamma_a}$ on any bounded interval $(a,b)$ and any $\gamma_a>-2$ as well as $V(x)=(1-x)/x$ on $(0,1)$.

If allowing also $\gamma_a=+\infty=\gamma_b$ (mimicking a potential with compact support) in the conjecture, we would get in this case (\ref{Conjecture}) with the constant term $-1/2$. Following the proof of the corollary above this would lead to (\ref{Weyl}) with $\vert\varepsilon(\lambda)\vert\leq1/2+\mathcal{O}_{\lambda\to\infty}(1/\lambda)$ for these potentials which is arguably the optimal error when approximating $N(\lambda)$.

\section*{Acknowledgements}
This work was supported in parts by the VILLUM Foundation grant no. 10059 and the grant 0135-00166B from the Independent
Research Fund Denmark. I thank Peter Hearnshaw and Martin Dam Larsen for their great interest in and helpful inputs to this work.

\printbibliography

\end{document}